\documentclass{iopart}
\usepackage{graphicx} % Include figure files
\usepackage{rotating}
\usepackage{cite}
\begin{document}

\title[Spectroscopy of Co$_2$Cr$_{1-x}$Fe$_{x}$Al.]
{Electronic structure and spectroscopy of the quaternary Heusler alloy Co$_2$Cr$_{1-x}$Fe$_{x}$Al }

\author{Sabine Wurmehl$^1$, Gerhard H. Fecher$^{1}$, Kristian Kroth$^{1}$,
        Florian Kronast$^2$, Hermann A. D\"urr$^2$,
        Yukiharu Takeda$^3$, Yuji Saitoh$^4$,
        Keisuke Kobayashi$^{3, 4}$,
        Hong-Ji Lin$^5$,        
        Gerd Sch\"onhense$^6$, and
        Claudia Felser$^1$}
\address{$^1$ Institut f\"ur Anorganische Chemie und Analytische Chemie, 
              Johannes Gutenberg - Universit\"at, D-55099 Mainz, Germany}
\address{$^2$ BESSY, D-12489 Berlin, Germany}
\address{$^3$ Synchrotron Radiation Research Center,
              Japan Atomic Energy Research Institute (SPring-8/JAERI), 
              Kouto 1-1-1, Mikaduki-cho, Sayou-gun, Hyogo, 379-5198, Japan}
\address{$^4$ Japan Synchrotron Radiation Research Institute (SPring-8/JASRI), 
              Kouto 1-1-1, Mikaduki-cho, Sayou-gun, Hyogo, 379-5198, Japan}
\address{$^5$ National Synchrotron Radiation Research Center - NSRRC, Hsinchu, 30076, Taiwan}
\address{$^6$ Institut f\"ur Physik, 
              Johannes Gutenberg - Universit\"at, D-55099 Mainz, Germany}

\ead{fecher@uni-mainz.de}

\date{\today}

\begin{abstract}
Quaternary Heusler alloys Co$_2$Cr$_{1-x}$Fe$_{x}$Al with varying Cr to Fe ratio $x$ were investigated experimentally and theoretically. The electronic structure and spectroscopic properties were calculated using the full relativistic Korringa-Kohn-Rostocker method with coherent potential approximation to account for the random distribution of Cr and Fe atoms as well as random disorder. Magnetic effects are included by the use of spin dependent potentials in the local spin density approximation.

Magnetic circular dichroism in X-ray absorption was measured at the $L_{2,3}$ edges of Co, Fe, and Cr of the pure compounds and the $x=0.4$ alloy in order to determine element specific magnetic moments. Calculations and measurements show an increase of the magnetic moments with increasing iron content. Resonant (560eV - 800eV) soft X-ray as well as high resolution - high energy ($\geq 3.5$keV) hard X-ray photo emission was used to probe the density of the occupied states in Co$_2$Cr$_{0.6}$Fe$_{0.4}$Al.
\end{abstract}

\pacs{75.50.Cc, 71.20.Lp, 78.70.Dm, 75.30.Cr}
\bigskip
\noindent{\it Keywords}: Heusler compounds, Electronic structure, Photo emission, Photo absorption
\submitto{\JPD}

\maketitle

\section{Introduction}

A great scientific interest is attracted by materials with a complete spin polarisation in the vicinity of the Fermi energy \cite{Pri98}. Such materials, being a metal for spin up and a semiconductor (or insulator) for spin down electrons, are called half-metallic ferromagnets \cite{GME83,CVB02} (HMF). Heusler compounds have been considered potential candidates to show this property \cite{GME83}. Theoretical calculations predicted an energy gap for minority electrons for the half-Heusler compound NiMnSb \cite{GME83, YM95} which, however, has been controversially discussed \cite{RNB00a,RNB00b,ZSV01}. Similarly, a HMF like behaviour was predicted for the Cobalt based full Heusler compounds Co$_2$MnZ (Z=Si,Ge) by Ishida \etal \cite{IMF98} and by Plogmann \etal \cite{PSB99} for Co$_2$MnSn.

Heusler compounds belong to a group of ternary intermetallics with the stoichiometric composition X$_2$YZ ordered in the $L2_1$ structure. X and Y are transition metals and Z is usually a main group element. Remarkably, the prototype Cu$_2$MnAl is a ferromagnet even though none of its constituents is one \cite{Heu03}.

Ferromagnetic properties of various Heusler compounds have been investigated experimentally \cite{WZ73,TFE74,EOS92,ZW74,EBE83,YYG01} and theoretically \cite{IAK82,PSB99} (see Refs.\cite{LB19C,LB32C} for comprehensive review). The Co-based Heusler compounds Co$_2$YZ are of particular interest because they show a comparatively high Curie temperature and varying magnetic moments ranging from $0.3 \mu_B$ to $1.0 \mu_B$ at the Co site depending on the constituents Y and Z (see Refs.: \cite{ZW74, Jez96, BNW00, YIA02}).

Co$_2$Cr$_{0.6}$Fe$_{0.4}$Al is of special interest because a relatively high magneto-resistance ratio of up to 30\% was found in powder samples in a small magnetic field of 0.1T \cite{BFJ03,FHK03}. Thin films of the compound were successfully grown by several groups \cite{KC04,HKO05a,HKO05b,JCB05}. A magneto-resistance ratio of 26.5\% \cite{IOG03} (at 5K) and 19\% \cite{ITO04} (at room temperature) was found for a tunnelling magneto-resistance (TMR) element of the same compound. Very recently, Marukame \etal \cite{MKM05} reported a TMR ratio of 74\% at 55K for a Co$_2$Cr$_{0.6}$Fe$_{0.4}$Al-MgO-CoFe magnetic tunnel junction. A spin polarisation of only less than 49\% was found for polycrystalline samples by means of Andreev reflections \cite{AJB03}. The observation of an incomplete spin polarisation may not only be caused by the model used to interpret the data \cite{AJB03,CFJ05} but also by the properties of the sample. Clifford \etal \cite{CVG04} reported recently a spin polarisation of 81\% in point contacts of Co$_2$Cr$_{0.6}$Fe$_{0.4}$Al.

X-ray photo absorption spectroscopy provides a way to study the unoccupied electronic structure of materials. In combination with circularly polarised light it allows the determination of the element specific magnetic moments. In particular, the L$_{2,3}$ edges of the ferromagnetic 3$d$ transition metals are easily accessible in the soft X-ray range (500eV to 800eV) as they exhibit pronounced white lines. X-ray magnetic circular dichroism (XMCD) \cite{SIL95,EFV03} was used in the present work to study the site specific magnetic moments in Co$_2$Cr$_{1-x}$Fe$_{x}$Al.

Photo emission spectroscopy is the method of choice to study the occupied electronic structure of materials. Excitation by low energies gives direct access to the band structure. However, low kinetic energies result in a low electron mean free path being only 6\AA at and below 100eV (all values calculated for Co$_2$CrAl using the TPP2M equations \cite{TPP93}). Thus, resonant excitation at the M edges of the 3$d$ metals may hardly be assigned as bulk sensitive because only one Heusler unit cell will contribute, for example. The situation becomes better at higher energies (18\AA at 700eV) close to the L absorption edges at of Fe and Co or for typical Al-K$_\alpha$ laboratory sources (31\AA at 1.4keV). In the hard X-ray region above 3.5keV one will reach true bulk sensitivity with an escape depth being larger than 65\AA for electrons emitted from the Fermi energy (128\AA at 8keV). High energy photo emission (at about 15keV excitation energy) was first performed as early as 1989 \cite{Mei89} using a $^{57}$Co M\"o\ss bauer $\gamma$-source for excitation. However, the resolution was very low due to the limitations given by the low intense source and single channel detection of the electrons \cite{KMG93}. Nowadays, high energy excitation and analysis become easily feasible due to the development of high intense sources (insertion devices at synchrotron facilities) and multi-channel electron detection. Thus, high resolution - high energy photo emission (HRHEPE) was recently introduced by several groups \cite{KYT03,SS04,TKC04,Kob05,PCC05,TSC05} as a bulk sensitive probe of the electronic structure in complex materials. In the present work, resonant photo emission (at the L$_{2,3}$-edges of the transition metals) and HRHEPE (3.5keV, 8keV) were used to study the density of states of Co$_2$Cr$_{0.6}$Fe$_{0.4}$Al.

The present work reports on the electronic, magnetic, and spectroscopic properties of the Heusler compound Co$_2$Cr$_{1-x}$Fe$_{x}$Al. The calculated properties are compared to experiments. Deviations from the $L2_1$ structure are discussed on hand of disordered, random alloys.

\section{Experiment}

The doped Heusler alloys Co$_2$Cr$_{1-x}$Fe$_{x}$Al were prepared by arc-melting under an argon atmosphere. The resulting specimens were dense polycrystalline ingots. Flat discs (8mm diameter by 1mm thickness) were cut from the ingots. The discs were mechanically polished for spectroscopic experiments. Remaining parts of the ingots were crushed to smaller size and partially pulverised for other investigations. The powders were made by hand in a mortar as steel ball milling resulted in a strong distortion of the structure as was proved by X-ray diffraction. The powder samples were used to allow for comparison of identical samples in spectroscopic and magneto-structural investigations. 

ESCA measurements excited by Al-K$_\alpha$ and Mg-K$_\alpha$ radiation were performed in order to check the composition and cleanliness of the samples. The structural properties of Co$_2$Cr$_{1-x}$Fe$_{x}$Al were determined by means of X-ray diffraction and extended X-ray absorption fine structure spectroscopy. The cubic structure with a lattice constant of about $5.73$\AA was confirmed for all samples under investigation.

Field dependent magnetic properties and saturation moments were measured by SQUID - magnetometry in the temperature range: $T=4\ldots300$K. All samples exhibited a soft magnetic behaviour. Magnetic circular dichroism (MCD) in X-ray absorption spectroscopy (XAS) was measured at the {\it First Dragon} beamline of NSRRC (Hsinchu, Taiwan) \cite{Che87,CS89}. The entrance and exit slits of the monochromator were set symmetrically in the range from 10$\mu$m to 25$\mu$m resulting in a resolution of about 50meV at the different absorption edges. Details of the data analysis are reported in \cite{EFV03}.

Soft X-ray excited photo emission was performed at the UE 56/1 PGM beamline of BESSY (Berlin, Germany). The resolution of the beamline was set to $E/\Delta E = 8000$ in the energy range from 600eV to 800eV. The spectra were taken by means of a hemispherical analyser (VG ESCA-Lab). The analyser slits were set to 200$\mu$m and a pass energy of 20eV was used. High resolution - high energy photo emission excited by hard X-rays \cite{KYT03} was performed at the BL22XU beamline of SPRING-8 (Hyogo, Japan) using a Gammdata Scienta R4000-10keV analyser \cite{TTN05}. For high resolution, the pass energy was set to 200eV and the entrance slit to 0.5mm. A 1.5mm slit was used for spectra with lower resolution, in particularly at 8keV photon energy were the valence band emission has a very low cross-section.

Before taking the spectra excited by synchrotron radiation, the samples were scratched in situ with a diamond file in order to remove the native oxide layer from the surface. ESCA and Auger electron depth profiling revealed a thickness of the oxide layer of up to 0.5$\mu$m depending on the sample history.

\section{Calculational Details}

Self-consistent band structure calculations were carried out using the spin polarised, full relativistic Korringa-Kohn-Rostocker method (SPRKKR) provided by Ebert \etal \cite{Ebe99,EMP01}. The exchange correlation functional was taken within the parametrisation of Vosko \etal \cite{WV77,VW80}. Random (or disordered) alloys were calculated within the coherent potential approximation (CPA). All Brillouin zone integrations were performed on base of a $22\times22\times22$ mesh of $k$-points. An imaginary part of 0.001Ry was added to the energy when calculating the density of states. 

The properties of the $L2_1$ compounds and random alloys where calculated in $F\:m\overline{3}m$ symmetry using the experimental lattice parameter ($a=5.727$\AA) as determined by X-ray powder diffraction. Co and Al where placed on 8c (1/4,1/4,1/4) and 4a (0,0,0) sites, respectively. Disorder between chromium and iron was accounted by setting the 4b (1/2,1/2,1/2) site occupations to $1-x$ for Cr and $x$ for Fe.

Several types of disorder have been considered. It turned out that the mostly occurring type of disorder is $B2$ like. This type is a result of disorder in the YZ (Y=Cr, Fe, Z=Al) planes. It is found by setting the $F\:m\overline{3}m$ site occupations of the 4a and 4b sites equally to 1/2 for Y and Z (assigned as $B2a$ in \cite{BP71}). The resulting structure with reduced $P\:m\overline{3}m$ symmetry was used for the calculations. The lattice parameter is $a_{B2}=a_{L2_1}/2$. X was placed at the 1a site (origin of the cube) and the site occupations for Y and Z at 1b (centre of the cube) were set equally to 1/2. For the mixed Fe-Cr systems, the site occupation factors for Y and Y' have to be weighted by $x$ and $(1-x)$, respectively. Complete disorder between all sites results in the $A2$ structure with reduced symmetry $I\:m\overline{3}m$.

\section{Magnetic and electronic properties of Co$_2$Cr$_{1-x}$Fe$_{x}$Al.}

The electronic structure of the pure and doped alloys will be discussed in the following. First, the electronic structure of the ordered alloys are presented, followed by the results for disordered alloys. Both include a more specific discussion of the magnetic properties on hand of measured and calculated magnetic moments. A band structure in the usual definition has no meaning for alloys with random disorder, due to the lack of periodicity. Therefore, only the integrated quantities will be discussed here. Indeed, the magnetic moments depend more directly on the DOS rather than on the particular form of the dispersion of the electronic bands.

\subsection{Electronic and magnetic structure of $L2_1$ ordered alloys.}

\begin{figure}
\includegraphics[width=8cm]{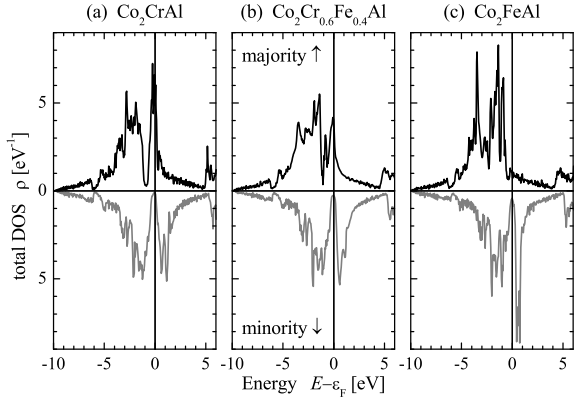}
\caption{ Density of states of Co$_2$Cr$_{1-x}$Fe$_{x}$Al ($x=0,0.4,1$).\newline
          (Note that the minority states are shown on a negative scale.) }
\label{fig1_dosall}
\end{figure}

First, the electronic structure for the $L2_1$ ordered ternary compounds was calculated. The resulting density of states (DOS) of Co$_2$CrAl and Co$_2$FeAl is shown in Fig.\ref{fig1_dosall}(a,c). These are the end members of the quaternary series of alloys. In the next step the electronic structure for the $L2_1$ ordered alloys Co$_2$Cr$_{1-x}$Fe$_{x}$Al was calculated and the result for $x=0.4$ is compared exemplarily in Fig.\ref{fig1_dosall}(b) to the pure compounds.

%%%%%%%%%%%%%%%%%%%%%%%%%%%%%
%%%%%%%%%%%%%%%%%%%%%%%%%%%%%

The low lying $sp$-bands are visible at about 6eV to 10eV below $\epsilon_F$. The $d$-bands start at about 5.5eV below $\epsilon_F$. The gap between $s$- and $d$-bands is less pronounced compared to non relativistic calculations (see Ref. \cite{FEF05}). The majority DOS at the Fermi energy decreases with increasing iron concentration $x$. The density of majority electrons at $\epsilon_F$ is a crucial point for spectroscopic methods investigating the spin polarisation, like spin-resolved photo emission. The small, non vanishing density in the half-metallic gap of the minority states at $\epsilon_F$ emerges mainly from the imaginary part added to the energy when calculating the DOS by means of the Greens function. However, a coupling between majority and minority states is always expected in the full relativistic calculations thus that a pure spin up state is not possible in general. See in Mavropoulos \etal \cite{MGP04,MSZ04} for a discussion of the relativistic effects and their influence on the half-metallic gap.

Figure \ref{fig1_dosall} reveals that the Fermi energy ($\epsilon_F$) is pinned in the minimum of the density of the minority states. Further, it is seen that the majority density is shifted with increasing Fe concentration and thus with increasing number of valence electrons. Both together have finally the result that the magnetic moments increase with the number of valence electrons (see below). This behaviour, in particular the pinning of $\epsilon_F$ in a minimum of the minority density, is typical for Slater-Pauling like behaviour in the case of materials with localised magnetic moments \cite{Sla36a,Pau38,MWM84,FKW05}. From the Slater-Pauling rule for ferromagnetic alloys with localised moments, the magnetic moment per atom is approximately given by $m=n_\uparrow - n_\downarrow = n_V- 2n_\downarrow \approx n_V-6$. Here, $n_V=n_\uparrow + n_\downarrow$ is the mean number of valence electrons per atom and given by the sum of the accompanied majority ($n_\uparrow$) and minority ($n_\downarrow$) electrons.  

\begin{figure}
\includegraphics[width=8cm]{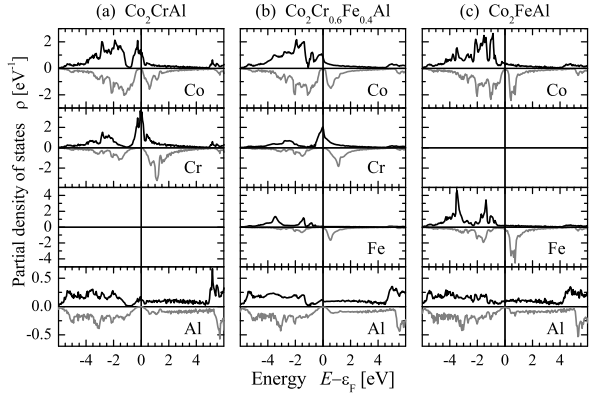}
\caption{ Spin resolved partial density of states of Co$_2$Cr$_{1-x}$Fe$_{x}$Al. \newline
         (Note the different scales at the partial DOS, the minority states are shown on a negative scale.) }
\label{fig2_dospart}
\end{figure}

The spin resolved partial density of states (PDOS) is displayed in Fig.\ref{fig2_dospart}. Shown are the partial (site specific) densities for the majority and minority states of Co$_2$Cr$_{1-x}$Fe$_{x}$Al for $x=0,0.4,1$. In particular the gap in the minority DOS is better resolved. From the behaviour of the different PDOS depending on the Fe concentration $x$, it is clear that the electronic structure does not follow a rigid band like model. In particular it is seen that the Cr PDOS decreases with increasing iron content keeping its shape rather unchanged. In the same way the Fe PDOS increases with $x$. However, the maxima of the density at the Co sites are clearly shifted away from $\epsilon_F$. The Al PDOS stays nearly unaffected and is rather independent of the Fe content.

%%%%%%%%%%%%%%%%%%%%%%%%%%%%%
%%%%%%%%%%%%%%%%%%%%%%%%%%%%%

From Fig.\ref{fig2_dospart}, it is seen that the high majority DOS at $\epsilon_F$ emerges from Cr. Both, Co and Fe exhibit only a small majority PDOS at and above $\epsilon_F$. Overall, the change of the majority DOS of Co$_2$Cr$_{1-x}$Fe$_{x}$Al around $\epsilon_F$ can be clearly attributed to the increasing amount of iron with respect to chromium. The minimum in the minority DOS around $\epsilon_F$ is mainly restricted by the shape of the Co PDOS. This indicates that the HMF like behaviour is mainly characterised by Co. The steep increase of the minority PDOS of Cr and Fe is mainly located in the unoccupied part above $\epsilon_F$. 

Doping with Fe changes not only the total DOS but also the PDOS of Co and much less pronounced the one of Cr. In particular, a very small shift of the Cr PDOS causes an additional decrease of majority states at $\epsilon_F$. This shift increases with increasing Fe concentration (as was found from the PDOS with variation of $x$, not shown here). The energy shift of the PDOS results in a change of the local magnetic moments in particular at the Co sites.

\begin{figure}
\includegraphics[width=8cm]{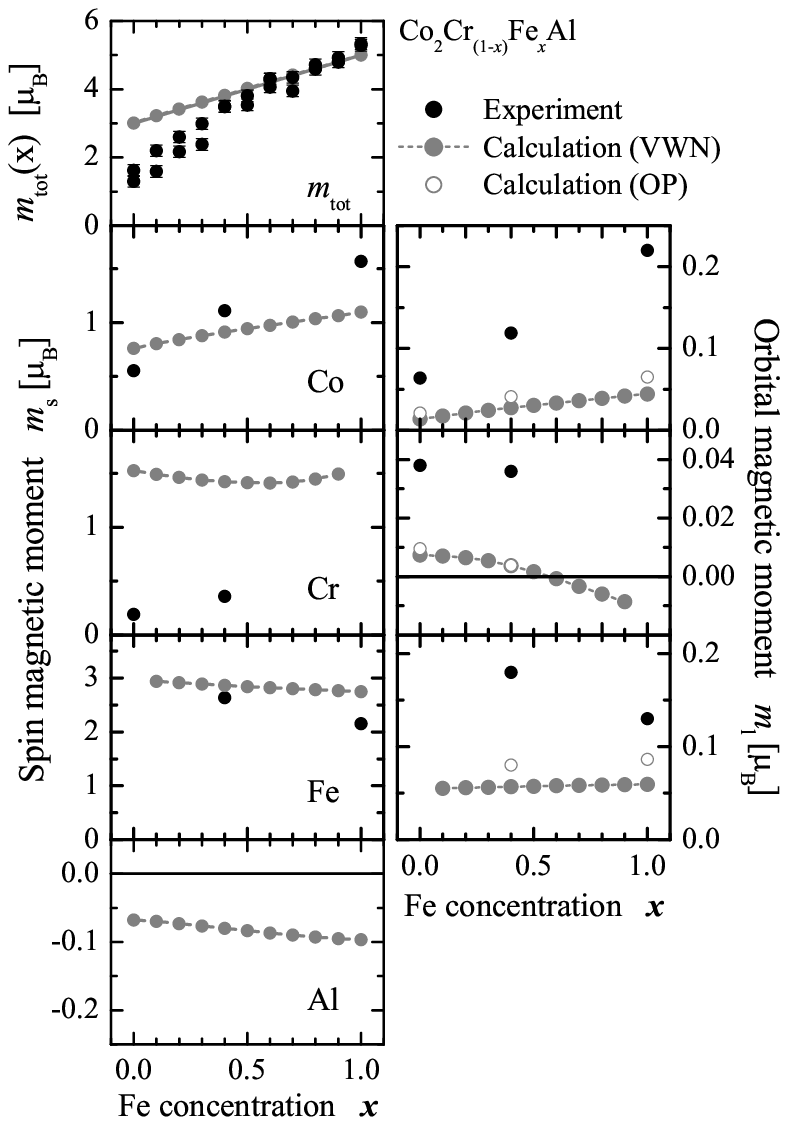}
\caption{ Magnetic moments in Co$_2$Cr$_{1-x}$Fe$_{x}$Al.\\
          The experimental uncertainty for a single measurement is within the width of the symbols as
          assigned at $m_{tot}$.}
\label{Fig_magmom}
\end{figure}

Figure~\ref{Fig_magmom} compares the measured and the calculated magnetic moments. The calculated total magnetic moment $m_{tot}=m_s+m_l$ depends linearly on the composition $x$ and follows the Slater-Pauling rule \cite{Kue84,MWM84,FKW05}. The measured total moments as determined by SQUID magnetometry at 5K are smaller compared to the calculated values in particular for low Fe (or equivalently high Cr) content. This behaviour will be discussed in the next section about disorder.

%%%%%%%%%%%%%%%%%%%%%%%%%%%%%
%%%%%%%%%%%%%%%%%%%%%%%%%%%%%

Site specific spin ($m_s$) and orbital ($m_l$) magnetic moments were determined for $x=0,0.4,1$ from XMCD measurements at the L$_{3,2}$ edges of Cr, Fe, and Co. Details of the spectroscopy experiments are reported below. The values were found from a sum rule analysis neglecting the magnetic dipole term \cite{EFV03}. The calculated spin magnetic moments at the Co sites increase with increasing Fe concentration $x$ as result of the shift of the majority PDOS of Co. The calculated spin moments at the Fe sites decrease slightly with increasing $x$. The principal dependence of the measured Co and Fe spin moments on $x$ is the same as in the calculation. However, the changes in the measured moments are stronger. The calculated Al spin moment is negative and very small. It is induced by the surrounding polarised Co atoms. Most obvious, the measured spin magnetic moments at the Cr sites are smaller compared to the calculated ones. This causes finally the too small total moment observed by SQUID magnetometry. The calculated orbital magnetic moments increase with $x$ at the Co sites by a factor of 4 and are nearly constant at Fe sites. The calculated Cr orbital moments are very small and change sign at $x=1/2$. In contrast, the measured orbital moments are throughout larger. Both experiment and calculation exhibit, however, the same trend for $m_l$.

Most evidently the calculated ratios between orbital $m_l$ and spin $m_s$ magnetic moments are much smaller than the experimental values. For Cr, a vanishing of the orbital momentum is expected, as observed in the calculation. On the other hand, the determination of the Cr moments at the $L_{2,3}$ edges is complicated due to the partial overlap of the lines, what may cause the observation (see spectra shown below in Fig.\ref{fig_xas}). The too low values of $m_l$ at Co and Fe sites cannot be explained that way. However, it is clear that the usual Hamiltonian used in LSDA contains directly a spin dependent coupling only. The orbital part of the moment thus arises only via spin-orbit interaction, or in words of the full relativistic case from the coupling of the large and the small components of the wave functions. This leads already in the case of pure Co and Fe metal to an underestimation of the orbital moments. One way to overcome that problem is the inclusion of the Brooks orbital polarisation \cite{EJB89} term (OP) in the Hamiltonian \cite{EB96}. The use of the OP term increases the orbital moments slightly as seen from Fig.\ref{Fig_magmom}, but evidently, the calculated values of the orbital magnetic moments are still smaller by a factor of up to 5 compared to the experimental values. The spin and total moments stayed in the same order as in pure LSDA. The observation of high orbital moments may point on additional orbital polarisation effects or on-site correlation being not respected in LSDA.

The findings of the full relativistic calculations are throughout compatible to those made in non-relativistic calculations using ordered compounds \cite{FKW05b}. Comparing the random alloy Co$_2$Cr$_{0.6}$Fe$_{0.4}$Al and the nearly iso-electronic ordered compound Co$_4$CrFeAl$_2$ \cite{FKW05b}, it is found that there are no major differences between CPA and virtual crystal approximation calculations for the materials investigated here. In particular, the rather integrated quantities like magnetic moments are the same. The non rigid band like character of the electronic structure upon Fe doping is revealed in both methods. The band gaps in the minority densities are here smaller due to the additional splitting of bands caused by the spin-orbit interaction rather than by properties of the CPA scheme, as is seen from the pure compounds.

\subsection{Random alloys with disorder.}

Table \ref{Tab:1} summarises the results of the calculations for the magnetic moment in the well ordered ($L2_1$), partially disordered ($B2$), and completely disordered ($A2$) alloy Co$_2$Cr$_{1-x}$Fe$_x$Al and compares them to the experimental values.

% Table 1 %%%%%%%%%%%%%%%%%%%%%%%%%%%%%%%%%%%%%%%%%%%%%%%%%%%
\begin{table}
\centering
\caption{Magnetic moments in Co$_2$Cr$_{1-x}$Fe$_x$Al.\\
    Given are the site specific moments per atom and the overall magnetic moment per unit cell, both
    in multiples of the Bohr magneton ($\mu_B$). The overall magnetic moments for the disordered structures
    are converted to fit those of the $L2_1$ cell in number of atoms. }
    \bigskip
    \begin{tabular}{c|ccc|cccc|ccc}
                  & $x=0$ &       &       & $x=0.4$ &       &       &         & $x=1$ &      &     \\
      Structure   & Co    & Cr    & tot   & Co      & Cr    & Fe    & tot     & Co    & Fe   & tot \\
      \noalign{\smallskip}\hline\noalign{\smallskip}
      $L2_1$      & 0.77  & 1.53  & 3.01  & 0.94    & 1.42  & 2.92  & 3.82    & 1.14  & 2.81 & 5.00 \\
      $B2$        & 0.75  & 1.46  & 2.92  & 0.57    & -1.12 & 2.95  & 1.61    & 1.07  & 2.91 & 4.95 \\
      $A2$        & 0.22  & 1.25  & 2.68  & 1.44    & 0.19  & 2.24  & 3.83    & 1.59  & 2.38 & 5.46 \\
      Exp.        & 0.55  & 0.19  & 1.3   & 1.11    & 0.36  & 2.64  & 3.49    & 1.57  & 2.15 & 5.29 \\      
      \noalign{\smallskip}\hline
    \end{tabular}
    \label{Tab:1}
\end{table}
%%%%%%%%%%%%%%%%%%%%%%%%%%%%%%%%%%%%%%%%%%%%%%%%%%%%%%%%%%%%%

Co$_2$CrAl and Co$_2$FeAl exhibit in $L2_1$ a total moment of $3.0 \mu_B$ and $5.0 \mu_B$, respectively. These values correspond to those expected from the Slater-Pauling rule for the moment in the HMF ground state.
The magnetic moment of correctly ordered Co$_2$Cr$_{0.6}$Fe$_{0.4}$Al is close to the value of $3.8\mu_B$ expected for a HMF ground state.  From the calculations it is expected that the total magnetic moment of the $B2$ disordered compounds is lower compared to the one in the $L2_1$ structure. 

The calculations for complete $B2$ or $A2$ disorder are not able to explain the too low value of the magnetic moment found experimentally for Co$_2$CrAl. A reduction of the overall moment caused by an anti-ferromagnetic order of the Cr atoms could not be verified by the calculations for these structure types. Therefore, calculations were also performed for other type of disorder. In the $DO_3$-type disordered alloy the Co and Cr atoms in 8c and 4b positions of the $L2_1$ structure are mixed. The calculations revealed an overall moment of $2.0\mu_B$. The moment of the Co atoms at the two different sites are $0.86\mu_B$ and $1.51\mu_B$. The moments of the Cr atoms are aligned anti-parallel with respect to each other and amount to $-0.41\mu_B$ and $+0.44\mu_B$ at 8c and 4b sites, respectively. Exchange of only one of the Co atoms with the Cr atom leads to the $X$ structure \cite{BP71}. In this case the calculations revealed an anti-parallel alignment of the Cr moments ($-0.93\mu_B$) with respect to the Co moments ($0.8\mu_B$ and $1.11\mu_B$) with a resulting overall moment of only $0.95\mu_B$ in the cell. From this approach, a tendency to ferr{\it \bfseries i}magnetic order is concluded if Cr and Co atoms change sites resulting in a lower magnetic moment. It is worthwhile to note that disorder of this type closes the gap in the minority DOS and the half-metallic character is lost.

The measured overall magnetic moment of Co$_2$Cr$_{0.6}$Fe$_{0.4}$Al is smaller compared to the calculated one for the ordered compound. Inspecting the site resolved moments one finds that the Co moment is slightly enhanced whereas the Cr and Fe moments are lowered compared to the $L2_1$ calculation. Such type of behaviour is also observed in the calculation for $A2$ disorder, but less pronounced. Thus it is obvious to assume a partial disorder for this type of samples. It is interesting to note that the calculations revealed a ferr{\it \bfseries i}magnetic ground state for $B2$ disordered Co$_2$Cr$_{0.6}$Fe$_{0.4}$Al, indeed, in contrast to the experiment where the Cr moment was small but aligned parallel to the Co moment.

The results from a calculation for 20\% disordered systems are compared in Tab.\ref{Tab:2}. The total moment for (80\%$L2_1$+20\%$B2$) is 4.04$\mu_B$ and thus clearly above the experimental value. The Co moment is slightly lower compared to the ordered structure. The Cr moments are at 20\% $B2$ disorder still aligned parallel to the Co moments. For (80\%$L2_1$+20\%$A2$), the Co and Fe moments are larger in the 4a and 4b positions compared to the 8c position. Most evidently the Cr moment exhibits an anti-parallel alignment in the 8c position. The total magnetic moment in the cell is 3.96$\mu_B$ and thus larger than the experimentally observed value. Comparing to the $L2_1$ ordered material, the enhancement of the Co moments with a simultaneous decrease of the Fe moments gives clear advice on $A2$ disorder in the experimentally examined samples.

% Table 2 %%%%%%%%%%%%%%%%%%%%%%%%%%%%%%%%%%%%%%%%%%%%%%%%%%%
\begin{table}
\centering
\caption{Magnetic moments in disordered Co$_2$Cr$_{0.6}$Fe$_{0.4}$Al.\\
         All calculated values are in $\mu_B$ per atom. The mean value is found using
         the disorder site occupation factors.}
    \bigskip
    \begin{tabular}{l c|ccc|c}
               &       & 8c    & 4a     & 4b    & mean \\
      \noalign{\smallskip}\hline\noalign{\smallskip}
               & Co    &  0.99 &  -     & -     & 0.99 \\
      20\%$B2$ & Cr    &  -    &  1.65  & 1.43  & 1.61 \\
               & Fe    &  -    &  2.88  & 2.93  & 2.89 \\      
            \noalign{\smallskip}\hline\noalign{\smallskip}
               & Co    &  1.08 &  1.86  & 1.98  & 1.25 \\
      20\%$A2$ & Cr    & -0.9  &  1.33  & 1.2   & 0.87 \\
               & Fe    &  0.96 &  2.79  & 2.87  & 2.43 \\      
      \noalign{\smallskip}\hline
    \end{tabular}
    \label{Tab:2}
\end{table}
%%%%%%%%%%%%%%%%%%%%%%%%%%%%%%%%%%%%%%%%%%%%%%%%%%%%%%%%%%%%%

In the experiment at Co$_2$FeAl, a larger total magnetic moment was observed in comparison to the expectation from the Slater-Pauling rule. At the same time the Co moment was enhanced and the Fe moment lowered compared to the calculations for $L2_1$. Comparing the values calculated for different type of disorder, one finds easily that the experimental values coincide with those expected for $A2$ type disorder. This means that the sample investigated by XMCD was to a large amount completely disordered.

\section{Spectroscopic properties of Co$_2$Cr$_{0.6}$Fe$_{0.4}$Al.}

In the following, results from spectroscopic experiments will be presented and compared to the calculated properties. The discussion in the two subsections about photo absorption and photo emission will be focused on the mixed compound Co$_2$Cr$_{0.6}$Fe$_{0.4}$Al.

\subsection{Photo absorption and XMCD.}

\begin{figure}
\includegraphics[width=8cm]{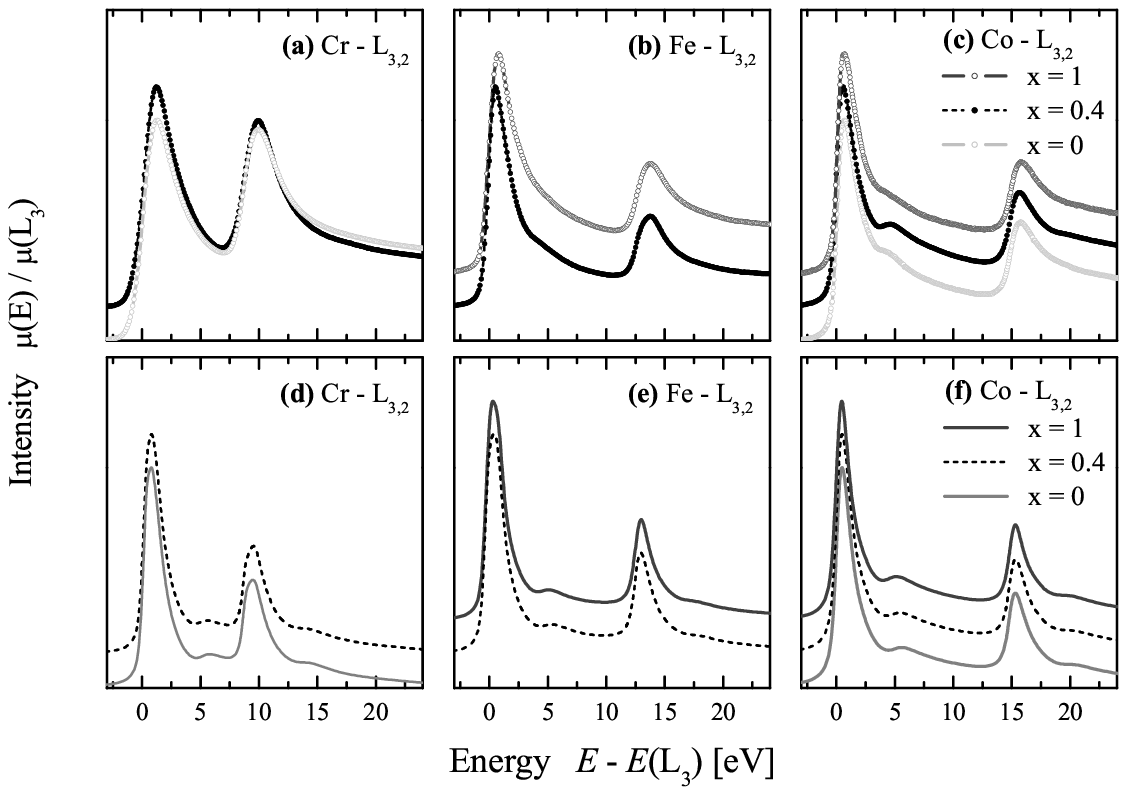}
\caption{ X-ray absorption spectra at the L$_{3,2}$ edges of Cr, Fe, and Co in Co$_2$Cr$_{1-x}$Fe$_{x}$Al.\newline
          (a) to (c) show the experimental and (d) to (f) the calculated spectra for $x=0,0.4,1$.
          The energy scales of the spectra are shifted with respect to the steepest point of the spectra
          (maximum of the derivative).
          The intensity scales are normalised to the maxima at the L$_3$ edges and shifted for sake of clarity.}
\label{fig_xas}
\end{figure}

X-ray absorption spectroscopy probes the unoccupied density of states above the Fermi energy. For the resonant excitation of the 2$p$-states of the 3$d$ transition elements the transition probability is proportional to the density of final states resulting from the 3$d$ and 4$s$ holes. The spectra, consisting mainly of the L$_3$ (2$p_{3/2}$) and L$_2$ (2$p_{1/2}$) white lines (see Fig.~\ref{fig_xas}), reflect the high density of unoccupied states resulting from the 3$d$ electrons.

%%%%%%%%%%%%%%%%%%%%%%%%%%%%%
%%%%%%%%%%%%%%%%%%%%%%%%%%%%%

Figure \ref{fig_xas} compares measured and calculated absorption spectra at the L$_{3,2}$ edges of Cr, Fe, and Co in Co$_2$Cr$_{1-x}$Fe$_{x}$Al. The monochromator resolution was set to 50meV for the spectra in (a)-(c). The spectra (d)-(f) were calculated for $L2_1$ structures using a lifetime broadening of 136meV. The measured white lines at both Cr edges are considerably wider compared to the calculated spectra whereas the width of the Fe and Co white lines is comparable. This points on a shorter lifetime of the holes at the Cr sites. Similarly, the measured L$_2$ lines of Co and Fe hint on a slightly larger lifetime broadening compared to the L$_3$ edges. Overall it is found that the spectra are governed by the lifetime broadening rather than by the experimental resolution. 

The experimental spectrum exhibits a prominent feature at the Co L$_3$ line for Co$_2$Cr$_{0.6}$Fe$_{0.4}$Al, in particular, that is shifted by about 4eV with respect to the white line. This feature is only weakly revealed for $x=0$ and $x=1$ and appears only as weak shoulder in the Fe ($x=0.4$) spectrum. It is not revealed in the measured Cr spectra what may be partially attributed to the higher lifetime broadening. The same structure occurs in the simulated absorption spectra and can be related to the structural properties. A similar feature is observed for example in fcc Ni but not in bcc Fe or hcp Co. It reflects the onset of the high lying $sd$-bands (see also Figs.\ref{fig1_dosall},\ref{fig2_dospart}). Here, its occurrence is characteristic for the highly ordered Heusler compounds. It vanishes for annealed and presumably disordered samples \cite{EFV03}. However, no particular features giving advice on half-metallic ferromagnetism are found in the absorption spectra what may partially be due to the relatively large width of the lines.

\begin{figure}
\includegraphics[width=8cm]{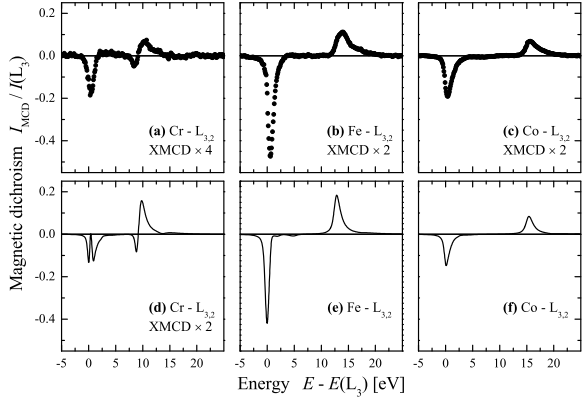}
\caption{ XMCD spectra at the $L_{3,2}$ edges of Cr, Fe, and Co in Co$_2$Cr$_{0.6}$Fe$_{0.4}$Al.\newline
          (a) to (c) show the experimental and (d) to (f) the calculated spectra.
          The XMCD signal  $I_{MCD}=\mu^{+}-\mu^{-}$ is normalised by the maximum intensity at the L$_3$ edges. 
          For better comparison, the experimental values are multiplied by a factor 2
          and both signals at the Cr edges are multiplied by an additional factor 2.}
\label{fig_mcd60}
\end{figure}

Figure \ref{fig_mcd60} compares measured and calculated XMCD spectra for Cr, Fe, and Co in Co$_2$Cr$_{0.6}$Fe$_{0.4}$Al. The experimental spectra shown in (a)-(c) were taken at 300K by switching a constant induction field of $B_0=\pm0.73$T (for field dependent measurements see \cite{EWF04}). The angle of photon incidence was 70$^o$ with a degree of circular polarisation of 85\%. The spectra (d)-(f) are calculated for parallel and anti-parallel orientation between magnetisation and photon spin.

%%%%%%%%%%%%%%%%%%%%%%%%%%%%%
%%%%%%%%%%%%%%%%%%%%%%%%%%%%%

Size and shape of the XMCD agree well at the Fe and Co edges comparing experiment (b,c) and calculation (e,f). It is clearly seen from the spectra at the Co edges (c,f) that the structural feature does not contribute to the XMCD signal. The calculated Cr XMCD (d) exhibits a splitting at the L$_3$ edge that is not resolved in the experimental spectra. However, the change of sign close to the L$_2$ edge is clearly visible in experimental as well as theoretical spectra. Both, splitting and change of sign, are due to the high density of majority states directly at the Fermi energy and the energetically shifted high density of the minority $d$-states above the half-metallic gap (compare Fig.\ref{fig2_dospart}).

\begin{figure}
\includegraphics[width=8cm]{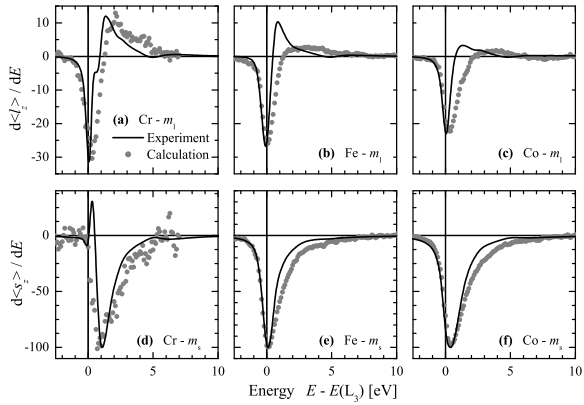}
\caption{ Orbital and spin polarisation in Co$_2$Cr$_{0.6}$Fe$_{0.4}$Al.\newline
          The measured dependencies were scaled to the minima for better comparison with the calculated data.}
\label{fig_compmcd}
\end{figure}

Figure \ref{fig_compmcd} shows the energy dependence of the spin ($s_z$) and orbital ($l_z$) densities (polarisation) in Co$_2$Cr$_{0.6}$Fe$_{0.4}$Al as derived from the XMCD and absorption spectra using the sum rules \cite{Ebe99}:
\begin{eqnarray}
	  {\frac{d}{dE}\left\langle s_z \right\rangle + 7 \frac{d}{dE}\left\langle T_z \right\rangle}
	  & \propto 3 (\Delta\mu_{L_3} - 2 \Delta\mu_{L_2}), \\
    {\frac{d}{dE} \left\langle l_z \right\rangle} & \propto 2 {(\Delta\mu_{L_3} + \Delta\mu_{L_2}).}
\label{eq1}
\end{eqnarray}
$T_z$ is the $z$ component of the magnetic dipole operator. $\Delta\mu_{L_{3,2}}$ assigns the difference of the absorption signal ($\mu$) for parallel and anti-parallel alignment of photon spin and magnetisation at the L$_3$ and L$_2$ edges.

% Table 3 %%%%%%%%%%%%%%%%%%%%%%%%%%%%%%%%%%%%%%%%%%%%%%%%%%%
\begin{table}
\centering
\caption{Magnetic moments and dipole operator in Co$_2$Cr$_{0.6}$Fe$_{0.4}$Al.\\
         (All calculated values are in $\mu_B$ per atom, The last column gives the number
         of occupied $d$-states per atom that are necessary to calculate the magnetic moments
         from the sum rule analysis.)}
    \bigskip
    \begin{tabular}{ c|ccc|c}
             & $m_s$ &  $m_l$ & 7 $T_z$ & $n_d$ \\
      \noalign{\smallskip}\hline\noalign{\smallskip}
       Co    & 0.86  & 0.042  &  0.0063 & 7.62 \\
       Cr    & 1.47  & 0.005  &  0.0009 & 4.68 \\
       Fe    & 2.59  & 0.082  &  0.0018 & 6.51 \\ 
       Al    & -0.07 & 0      &  0      & 0.4  \\             
      \noalign{\smallskip}\hline
      Co$_2$Cr$_{0.6}$Fe$_{0.4}$Al & 3.56 & 0.12 & - & 21\\
    \end{tabular}
    \label{Tab:3}
\end{table}
%%%%%%%%%%%%%%%%%%%%%%%%%%%%%%%%%%%%%%%%%%%%%%%%%%%%%%%%%%%%%

The calculated values for the magnetic dipole operator are given in Tab.\ref{Tab:3} and compared to the magnetic moments. All values in Tab.\ref{Tab:3} were calculated using the OP scheme. The magnetic dipole term $7T_z$ is in all cases small compared to the spin magnetic moment and thus can safely be neglected in the data analysis. 

%%%%%%%%%%%%%%%%%%%%%%%%%%%%%
%%%%%%%%%%%%%%%%%%%%%%%%%%%%%

The experimental polarisation spectra are compared to calculated spectra in Fig.~\ref{fig_compmcd}. The Cr XMCD spectrum is particularly interesting because in the disordered areas of the sample the Cr moments may be aligned partially anti-parallel to each other and the magnetic signal cancels out. Therefore, the remaining Cr XMCD contains information from the ordered regions, only. While the agreement between theory and experiment is only moderately well for the XMCD (Fig.\ref{fig_mcd60}(a,d)), the spin and orbital moment densities (Fig.\ref{fig_compmcd}(a,d)) are described remarkably well. The prominent maximum in the spin density at about 1eV above the $L_3$ edge is clearly revealed. The change of sign close to the Fermi edge is indicative for the large exchange splitting of the Cr 3$d$ states. The orbital densities exhibit a change of sign for all elements leading finally to the small orbital moment after integration. The experimental spectra are throughout wider than the calculated ones resulting in the higher experimental values for the orbital moments \cite{FEF05}. However, the width of the spin densities fits that of the calculated spectra.

\subsection{Photo emission.}

Using photo emission spectroscopy one obtains information about the occupied electronic states. The dispersion of the bands may be studied by spin and angular resolved UPS. This needs, however, well ordered surfaces of single crystals. For the interpretation one has to take into account that usually the kinetic energy of the photo emitted electrons is in a range where the mean free path is only a few \AA and the method therefore mainly shows electronic bands very close to the surface layer.

Using higher photon energy for excitation (and thus kinetic energies of the electrons), it is possible to study the density of states, in particular for polycrystalline samples. Making use of resonant excitation one expects to become more bulk sensitive \cite{SIM00} and may be able to distinguish the site specific contributions.

\begin{figure}
\includegraphics[width=7.5cm]{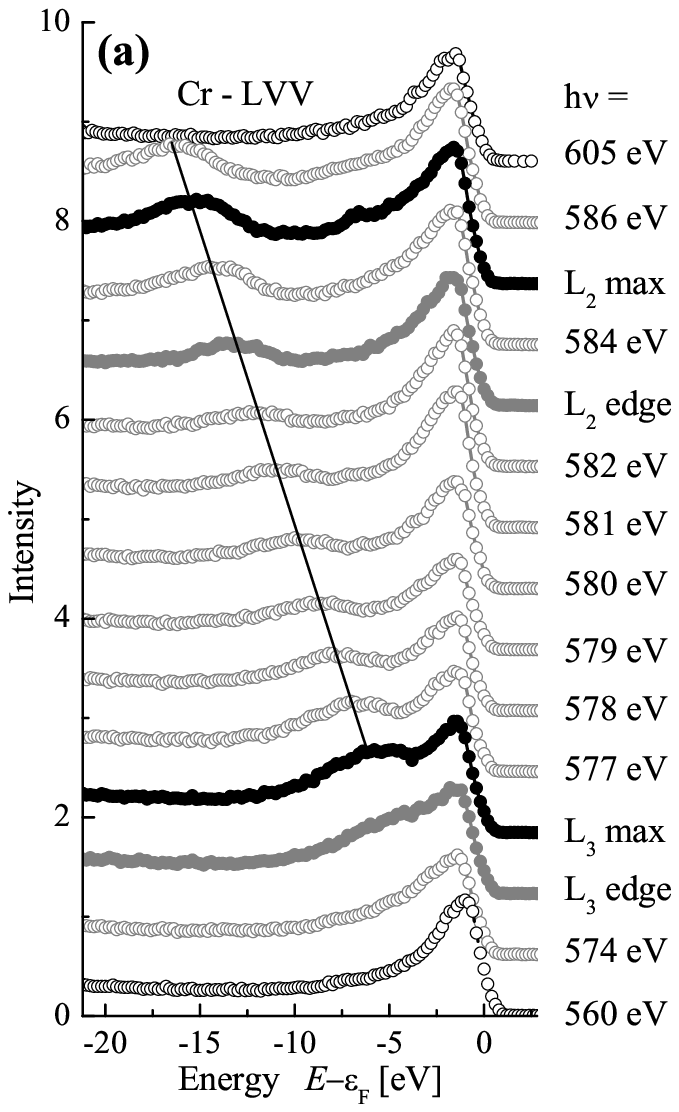}
\includegraphics[width=7.5cm]{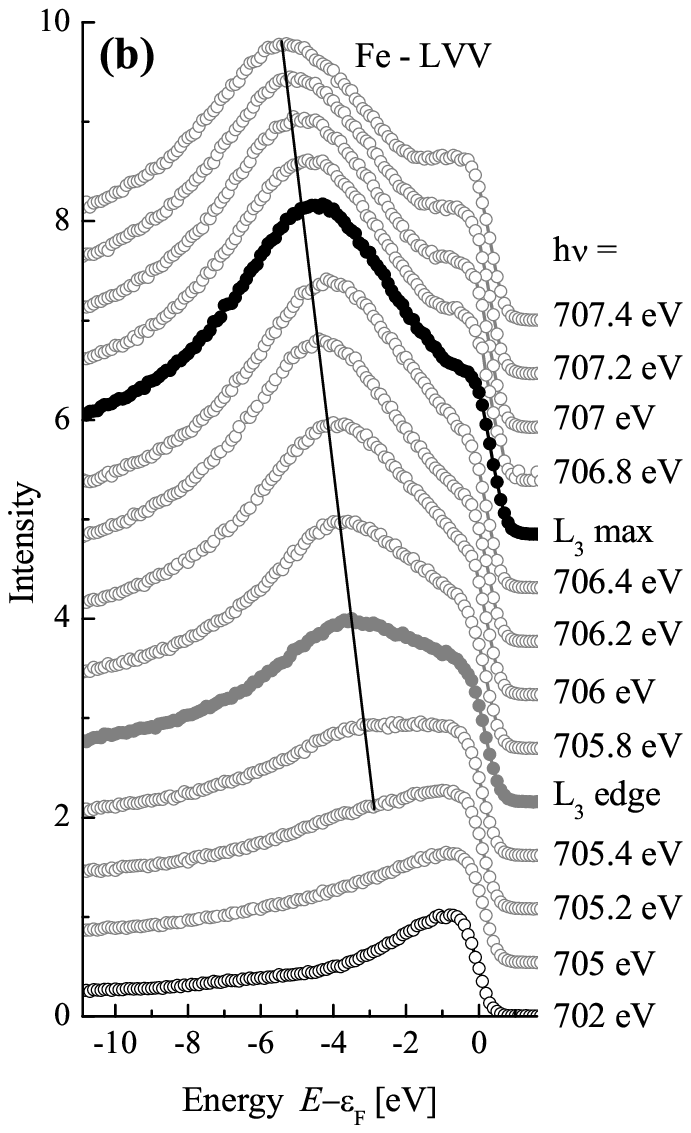}
\caption{ Valenceband-XPS of Co$_2$Cr$_{0.6}$Fe$_{0.4}$Al.\\
          Photo emission spectra excited by photon energies
          close to the Cr L$_{3,2}$ (a) and close to the Fe L$_3$ (b) absorption edges.
          The lines along the L$_3$VV Auger transitions are drawn to guide the eye. 
          Photon energies are un-calibrated monochromator readings and the onset and maxima of the white lines
          are marked.}
\label{fig_rescr}
\end{figure}

% Low Energy PE %%%%%%%%%%%%%%%%%%%%%%%%%%%%%%%%%%%%%%%%%%%%%%%%%%%%
Figure \ref{fig_rescr} shows energy dependence of the valence band photo emission spectra taken with photon energies close to the Cr L$_{3,2}$ and Fe L$_{3}$ edges. The spectra are normalised to the photon flux. The given photon energies correspond to the un-calibrated monochromator readings. Therefore, the onset and maxima of the white lines as found from the maximum of the energy derivatives of the absorption spectra are marked. The spectra excited close to the Cr L$_{3,2}$ edges show a 2eV wide feature close to the Fermi energy (see Fig.\ref{fig_rescr}(a)). No particular enhancement of the intensity is observed if the photon energy is crossing the Cr L$_3$ white line of the absorption spectrum. The L$_3$VV Auger electron emission is observed just when the onset of the Cr L$_3$ absorption edge is reached. It can be clearly identified from its linear energy dependence assigned by a straight line in Fig.\ref{fig_rescr}(a). On a kinetic energy scale it stays fixed. The off-resonant spectra (560eV, 605eV) are very similar in the valence band region (-10eV$\ldots$0eV). No particular features are detectable if crossing the L$_2$ edge.

%%%%%%%%%%%%%%%%%%%%%%%%%%%%%
%%%%%%%%%%%%%%%%%%%%%%%%%%%%%

Figure \ref{fig_rescr}(b) shows the valence band spectra taken with photon energies close to the Fe L$_3$ absorption edge. The intensities at the Fermi energy are enhanced by a factor of $\approx1.6$ in the photon energy range of 706eV to 707eV compared to excitation before the edge at 702eV. As for Cr, the evaluation of the L$_3$VV Auger electron emission is clearly visible.

The spectra observed here at the Cr and Fe L$_3$ resonance are obviously different from those reported by H\"ufner \etal \cite{HYM00} for elemental metals, even if accounting for the higher analyser resolution used in that work. (The present overall resolution was about 100meV.) In particular, the L$_3$VV Auger emission is less pronounced. An Auger signal with varying kinetic energy was not detectable. The differences in the resonant behaviour as well as the Auger emission are clearly related to the differences between the materials that is an alloy here compared to pure metals in \cite{HYM00}.

\begin{figure}
\includegraphics[width=7.5cm]{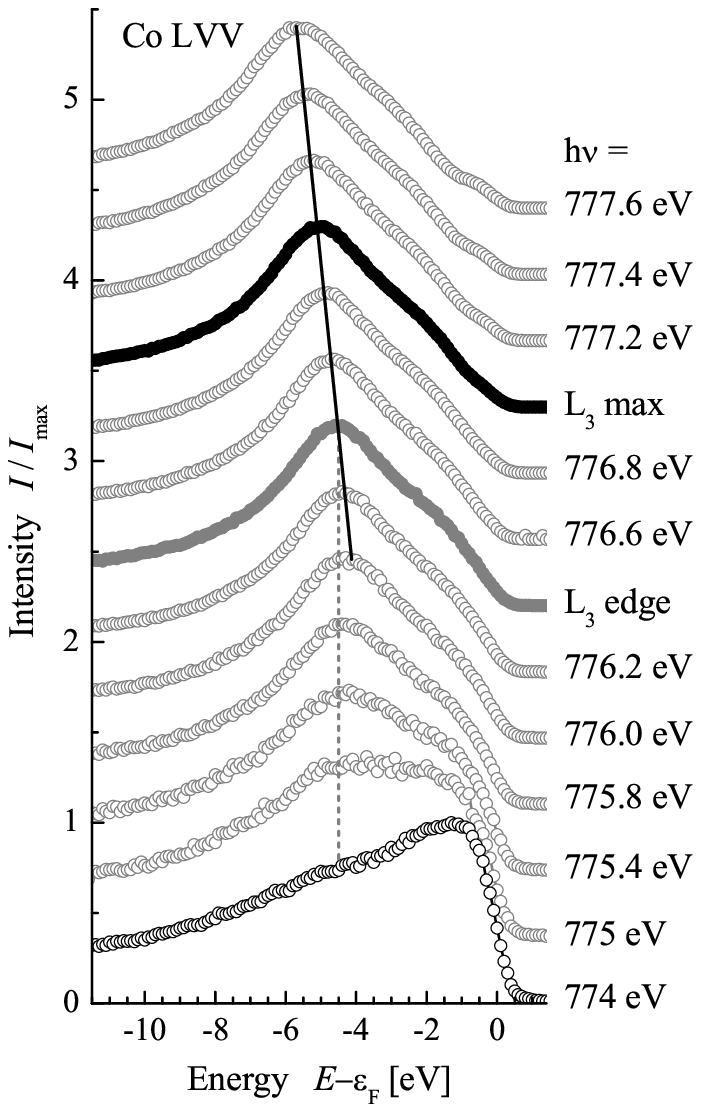}
\caption{ Resonant Co-valenceband-XPS of Co$_2$Cr$_{0.6}$Fe$_{0.4}$Al.\\
          Resonant photo emission spectra excited by photon energies
          close to the Co L$_{3}$ absorption edge.
          The energies corresponding to the onset and the maximum of the white line are marked.
          The spectra are normalised to the maximum intensity (see text). 
          The full line along the L$_3$VV Auger transition is drawn to guide the eye. 
          The dashed line assigns a dispersionless feature.}
\label{fig_resco}
\end{figure}

Figure \ref{fig_resco} shows the photon energy dependence of the valence band photo emission spectra taken with photon energies close to the Co L$_3$ edge. The spectra exhibited a strong variation of the intensity with the photon energy. Therefore, they were normalised to their maxima for better comparison. Above the onset of the L$_3$ white line, the intensity at $\epsilon_F$ is enhanced by a factor $5\ldots3.5$ with increasing photon energy. Again, the constant-kinetic-energy L$_3$VV Auger electron emission is clearly observed if the onset of the L$_3$ absorption edge is reached. Moreover, at energies before the edge an additional dispersionless feature (note the larger energy steps for lower photon energies) becomes visible at 4.5eV binding energy. Its maximum at 776.6eV excitation energy is about 35$\times$ more intense as the intensity at 774eV photon energy. Part of this very high intensity is, however, due to the intense underlying L$_3$VV Auger electron emission. This type of feature was previously assigned as two-hole satellite \cite{HYM00} with varying two-hole correlation energy. 

%%%%%%%%%%%%%%%%%%%%%%%%%%%%%
%%%%%%%%%%%%%%%%%%%%%%%%%%%%%

Other than at the Cr and Fe edges, a clear resonance like behaviour is observed in the valence band spectra if crossing the Co L$_3$ absorption edge, especially at the Fermi energy. This may be due to the particularly higher order of the Co atoms in the compound. Already if neglecting any B2 type disorder, Fe and Cr are randomly distributed on the regular sites of the L2$_1$ structure of the Heusler compounds leading to an alloy like behaviour. The small enhancement of intensity at the Fe edge presumably reflects a higher order of those atoms compared to Cr.

\begin{figure}
\includegraphics[width=10cm]{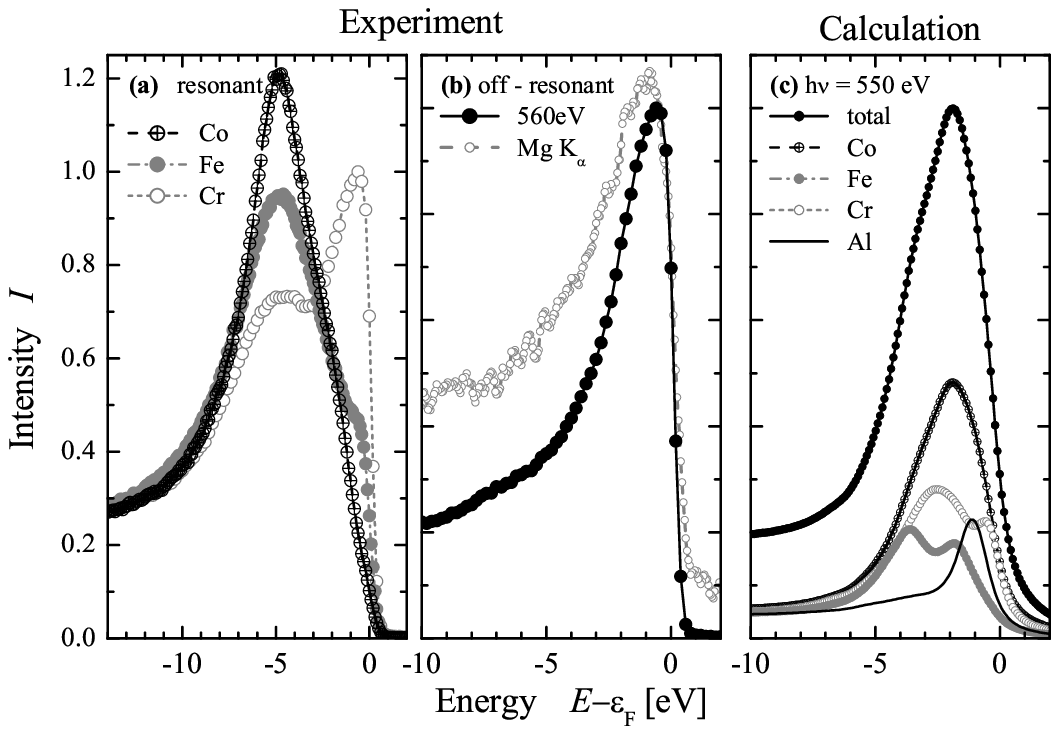}
\caption{ Soft X-ray excited XPS of Co$_2$Cr$_{0.6}$Fe$_{0.4}$Al.\\
          Shown are resonant (a) and off-resonant (b) photo emission spectra
          in comparison to calculated site-specific spectra (c) \cite{FEF05}. 
          (Note the slightly different energy scale in (a), for
          normalisation of the spectra see text.)}
\label{fig_rpecomp}
\end{figure}

Figure \ref{fig_rpecomp} compares measured and calculated photo emission spectra for polycrystalline Co$_2$Cr$_{0.6}$Fe$_{0.4}$Al. Shown are the resonantly excited spectra (a), the off-resonant spectra (b) and the calculated spectra (c). The on-resonance spectra are taken at the maxima of the L$_3$ absorption edges of Co, Cr, and Fe. They were normalised to the intensity at 15eV binding energy for better comparison. The off-resonance spectrum was taken at 560eV, that is well below the onset of the Cr L$_3$ edge. The spectrum excited by Mg K$_\alpha$ radiation is shown for comparison in Figure \ref{fig_rpecomp}(b). The calculated spectrum in Fig.~\ref{fig_rpecomp}(c) shows the total intensity for 550eV photon energy together with the site resolved spectra for Co, Cr, Fe, and Al. These spectra were calculated using a 0.1Ry Cauchy-Lorentz broadening without accounting for energy dependent hole lifetimes.

%%%%%%%%%%%%%%%%%%%%%%%%%%%%%
%%%%%%%%%%%%%%%%%%%%%%%%%%%%%

The resonantly excited spectra at the absorption edges of the elements in Co$_2$Cr$_{0.6}$Fe$_{0.4}$Al are governed by the L$_3$VV Auger lines and thus do not allow an element specific discrimination of the density of states, unfortunately. This is not a matter of resolution but explained by the high intrinsic width of the Auger lines. It is clearly seen that the ratio between the Auger line and the intensity at the Fermi energy depends strongly on the element. It is highest for Co ($>10$) but only very small for Cr ($<1$). Fe lies in between and the intense emission close to the Fermi energy is still visible. The off-resonance spectrum exhibits a sharp onset at the Fermi energy with a maximum of intensity at 0.7eV binding energy. The low lying $s$-bands contribute only weakly to the intensity at binding energies between 5eV and 10eV. The calculated spectrum exhibits a high intense peak at 1.9eV binding energy that can clearly be attributed to emission from states located at the Co sites. In comparison, a light enhancement of the intensities at energies between 3.5eV and 7eV below $\epsilon_F$ is observed in the spectrum excited by Mg K$_\alpha$ radiation. 

The discrepancy between the intensity maxima observed in the experiment compared to the calculated one may be explained as follows. The calculated spectrum is a true bulk spectrum with the surface neglected completely whereas the experimental spectrum at 550eV photon energy is still dominated by surface effects as the electron mean free path is in the order of 10\AA and thus less than the height of two cubic $L2_1$ cells. From the calculated spectrum one expects high emission from Cr and Al sites at 0.6eV and 1.0eV binding energy, respectively. Thus the comparison between experiment and calculation indicates that Cr and Al may have segregated to the surface to some extent, or equivalently, it points on a loss of Co and Fe in the region close to the surface. In addition, the measured high intensity close to the Fermi energy is partially due to the high lifetime of holes at $\epsilon_F$ that is not directly accounted in the calculation.

High energy photo emission was performed using monochromised X-rays of up to 8keV energy. The energy scale was calibrated at the Fermi energy of the Au valence band and $4f$ spectra (not shown here). The overall resolution (optics plus analyser) was determined from the Au valence band spectrum to be better than 55meV at 3.5keV under the same conditions used for the spectra discussed in the following. 

\begin{figure}
\includegraphics[width=8cm]{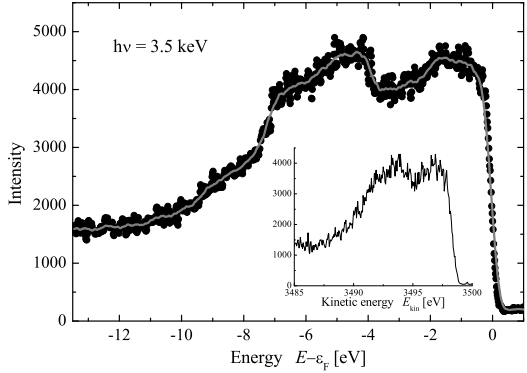}
\caption{ High resolution - high energy VB-XPS of Co$_2$Cr$_{0.6}$Fe$_{0.4}$Al.\\
          The photo emission spectra are excited by hard X-rays of 3.5keV energy.
          The inset shows for comparison a spectrum taken with lower resolution 
          plotted versus kinetic energy. }
\label{fig_hevb}
\end{figure}

\begin{figure}
\includegraphics[width=8cm]{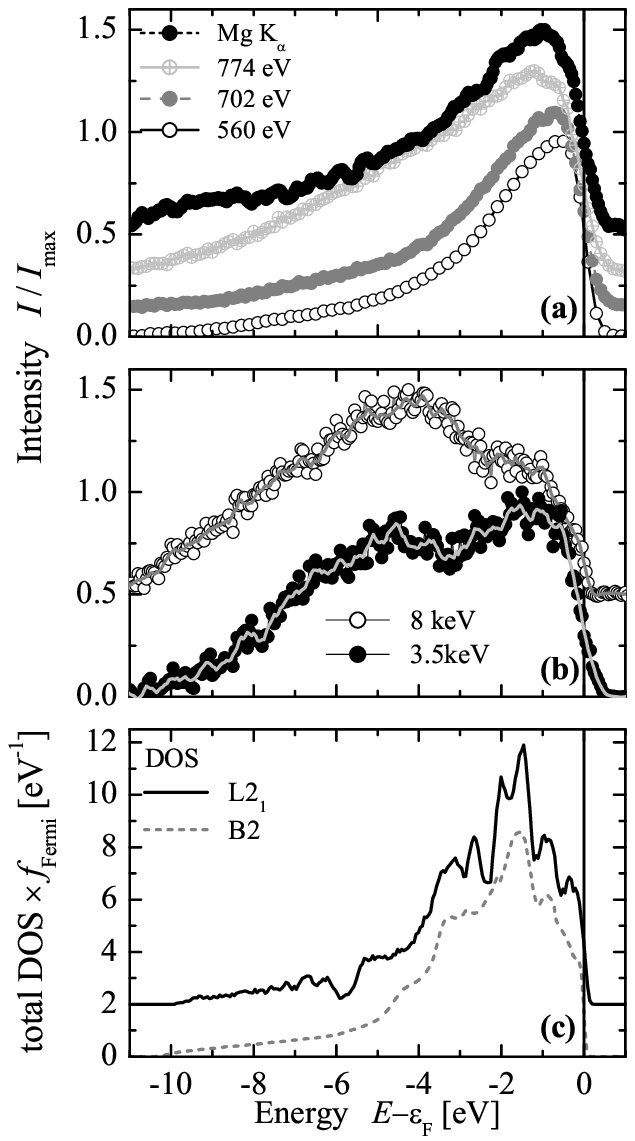}
\caption{ PE spectra and DOS\\
          Compared are the valence band spectra of Co$_2$Cr$_{0.6}$Fe$_{0.4}$Al excited by low (a) 
          and high (b) photon energies to the calculated, spin averaged, total density of states (c).
          (For applied data manipulation see text.)}
\label{fig_pecompare}
\end{figure}

Figure \ref{fig_hevb} shows the valence band spectrum of Co$_2$Cr$_{0.6}$Fe$_{0.4}$Al excited by hard X-rays. A broad feature of about 2eV width appears close to the Fermi energy. In addition, another broad feature becomes visible between 4eV and 7eV below the Fermi energy that is also revealed in the low resolution spectrum (see inset in Fig.\ref{fig_hevb}). Subtracting a Shirley type background (compare Fig.\ref{fig_pecompare}(b)) reveals an overall width of the valence band of about 10eV. Other than in the low energy spectra, there is no pronounced overshooting of the intensity close to the Fermi energy. This results in better agreement with the DOS showing its maximum at 1.5eV below $\epsilon_F$. The low energy spectra are governed by the high lifetime of the holes close to the Fermi energy. At high energy excitation (3.5keV) the electrons are faster by a factor of about 2.2 (compared to 700eV) thus the interaction of the outgoing electron with the remaining $N-1$ electron state of the solid becomes weaker. This observation gives clear evidence that one has to work at much higher energies in order to come close to the limit of sudden approximation.

%%%%%%%%%%%%%%%%%%%%%%%%%%%%%
%%%%%%%%%%%%%%%%%%%%%%%%%%%%%

% PE comparison %%%%%%%%%%%%%%%%%%%%%%%%%%%%%%%%%%%%%%%%%%%%%%%%%%%%
Figure \ref{fig_pecompare} compares the measured photo emission spectra for polycrystalline Co$_2$Cr$_{0.6}$Fe$_{0.4}$Al with the calculated total DOS. For better comparison, a Shirley type background was subtracted from all photo emission spectra and the spectra were normalised to their maxima. The low energy spectra (a) are shifted additionally in intensity by an offset of 0.15 with respect to each other. For better comparison, the high energy spectra (b) taken with a lower but equal resolution are shown and also shifted with respect to each other. The total density of states (c) was convoluted by a Fermi distribution for 300K after averaging over both spin components. The $L2_1$ curve is shifted by 2eV for better comparison. 

%%%%%%%%%%%%%%%%%%%%%%%%%%%%%
%%%%%%%%%%%%%%%%%%%%%%%%%%%%%

The DOS for the $L2_1$ structure can be distinguished in two regions: first, the low lying $sp$-band between 6eV and 10eV binding energy and second, the $d$-bands ranging from about 5.5eV up to the Fermi energy. The DOS for the $B2$ structure is similar, but the low lying gap between $sp$ and $d$ bands is not longer revealed. The off-resonant, low energy, spectra are governed in the excitation energy range from 560eV to 1.2keV by a high intensity close to the Fermi energy. The low lying $sp$-band or the Heusler gap between $sp$ and $d$-bands are hardly detectable in those spectra. The measured high resolution - high energy spectrum exhibits a width of the $d$-bands of about 7eV with a clear minimum at about 3eV binding energy as well as the low lying $sp$-band extended up to 10eV binding energy. The differences in the $d$-band emission and the total DOS are obvious. The maximum of the DOS at about 1.5eV below $\epsilon_F$ is less pronounced in the HE-spectra and absent at low excitation energies. Suppose that electron correlation can not be neglected even though the $d$-states are de-localised without doubt. In that case one will expect a splitting of the $d$-bands increasing its width that is not reproduced by pure LSDA calculations. It is worthwhile to note that the off-resonant spectra exhibit an enhancement at energies of approximately 4eV to 7eV below $\epsilon_F$ for excitation energies above 750eV. Both high energy spectra, taken at 3.5keV and 8keV, reveal a high intensity at about $(-4\ldots-7)$eV binding energy where the LSDA calculations predict the gap between the low lying $sp$ band and the $d$ bands. Moreover, the high intensity feature at about 4.5eV binding energy is not only observed in the high energy spectra. It corresponds roughly to the dispersionless feature revealed by the resonantly enhanced low energy spectra (seen at the Co L$_3$ edge). There is, however, no according high density observed in the DOS, neither in total nor partially for Co. From this point of view, the high energy spectra give evidence that electron-electron or electron-hole correlation may play an important role in Heusler compounds, at least for those being Co$_2$ based like the one investigated here. The new feature seen by bulk sensitive photo emission corresponds to a band-complex between 2.2eV and 5.5eV below $\epsilon_F$. It cannot be explained by disorder but may point on some deficiencies of the local spin density approximation.

Overall, the photo emission experiments show that the material has a high density of states close to the Fermi-energy. This has to be seen as an advantage of Co$_2$Cr$_{0.6}$Fe$_{0.4}$Al compared to other predicted half-metallic ferromagnets like the high moment compounds NiMnSb or Co$_2$MnSi. In the latter compounds with above $1\mu_B$ per atom, the density of majority states at $\epsilon_F$ is arising from few strongly dispersing $d$ bands crossing the Fermi energy. This results in a low majority density of states at $\epsilon_F$. Any even small perturbation of the minority gap in such systems will lead to a strong decrease of the spin polarisation and thus causes a strong reduction of effects in spin dependent transport properties. The large majority density at $\epsilon_F$ explains also why the spin polarisation stays rather high in partially $B2$ disordered Co$_2$Cr$_{1-x}$Fe$_{x}$Al in agreement to Miura \etal \cite{MNS04}.

In conclusion, it was shown that full relativistic SPRKKR calculations are very useful to explain spectroscopic experiments. The KKR method with CPA was successfully used to explain experimental spectra including random disorder in the investigated samples. The use of a full relativistic scheme allows the correct determination of transition matrix elements needed to calculate spectroscopic quantities, in particular if the photon or electron spin plays a role. 

\section{Summary}
In summary, results from spectroscopy experiments (XMCD, soft and hard X-ray valence band photo emission) at quaternary Heusler alloys Co$_2$Cr$_{1-x}$Fe$_{x}$Al were analysed in the light of ab-initio calculations. Special attention was focused on Co$_2$Cr$_{0.6}$Fe$_{0.4}$Al. The measured overall and site specific magnetic moments are in accordance with the calculations, if disorder is assumed. In particular, it was shown that Co-Cr disorder leads to the low magnetic moment observed in Co$_2$CrAl. It points on ferrimagnetic order with an anti-parallel alignment of part of the Cr moments in disordered alloys.

It was found by resonant and high energy photo emission that there is a discrepancy between the experimentally observed and the theoretically calculated density of states in Co$_2$Cr$_{0.6}$Fe$_{0.4}$Al. This observation suggests the presence of correlation in Heusler compounds being not accounted for by local spin density approximation in its current form. Moreover, strong differences in surface and bulk photo emission spectra (excited at photon energies from 500eV to 8keV) reveal the loss of the bulk signature if emission takes mainly place from the surface layer.

\section{Acknowledgments}

The authors are grateful to O.~Rader (BESSY, Berlin), H.-J.~Elmers, H.~C.~Kandpal, and U.~Stumm (Uni-Mainz) for support and help with the experiments and calculations. Further thanks go to all members of the synchrotron facilities NSRRC (Hsinchu, Taiwan), Spring-8 (Hyogo, Japan), and BESSY (Berlin, Germany) for their help and support during the beamtimes.

Financial support by the Deutsche Forschungs Gemeinschaft (Forschergruppe 559), DAAD (03/314973 and 03/23562) is gratefully acknowledged.

\end{document}